# Information entropy and thermal entropy: apples and oranges

L.B. Kish [1]  and  D.K. Ferry [2]

[1] *Department of Electrical and Computer Engineering, Texas A&M University, TAMUS 3128, College Station, TX 77843-3128*

[2] *School of Electrical, Computer, and Energy Engineering and Center for Solid State Electronic Research, Arizona State University, Tempe, AZ 85287-5706, USA*

**Abstract.**
The frequent misunderstanding of information entropy is pointed out. It is shown that, contrary to fortuitous situations and common beliefs, there is no general interrelation between the information entropy and the thermodynamical entropy. We point out that the change of information entropy during measurement is determined by the resolution of the measurement instrument and these changes do not have a general and clear-cut interrelation with the change of thermal entropy. Moreover, these changes can be separated in space and time. We show classical physical considerations and examples to support this conclusion. Finally, we show that information entropy can violate the Third Law of Thermodynamics which is another indication of major differences from thermal entropy.

**1 Introduction**

*1.1 Information entropy about physical systems*

When von Neumann asked Shannon how he was getting on with his information theory, Shannon replied as follows [1]:

"*The theory is in excellent shape, except that I need a good name for "missing information"*".

Von Neumann then suggested [1]:

"*Why don't you call it entropy. In the first place, a mathematical development very much like yours already exists in Boltzmann's statistical mechanics, and in the second place, no one understands entropy very well, so in any discussion you will be in a position of advantage.*"

Resonating with von Neumann's joke, since then, information entropy has not only been used but it has also been misused due to misunderstanding its meaning that has not been clarified in physical informatics. It is obvious that, in the physical information channel and during measurements, the existence of thermodynamic fluctuations make statistical physics enter into the information entropy, but is there more and, if yes, what exactly is that?

In this paper we point out that the two entropies have different physical nature. Thermodynamical entropy of a closed physical system is well-defined objective physical quantity. On the other hand, information entropy manifest via the interaction of the *measured physical systems* (PS) and the *measurement instrument* (MI). The MI is the detector and information channel between the PS and the observer. As a consequence, the information entropy is a *subjective*, *measurement-system-dependent* feature while the thermodynamical entropy is an *objective property* of the physical system. In Section 4, we will show experimental situations supporting this assessment.

In Section 2, we briefly define the two entropy types; in Section 3 we review the basics of minimum energy dissipation in binary measurement systems; in Section 4 we show two classical physical experimental schemes elucidating the differences between these entropy measures; in Section 5, we recall two simple proofs about the non-validity of the Landauer theorem; and in Section 6 we point out that the information entropy is able to violate the Third Law of Thermodynamics, which also proves that the two entropies are apples and oranges.

## 2 Thermal entropy and information entropy

*2.1 Thermal entropy*

Consider a closed physical system. The thermal entropy $S_t$ is a macroscopic state variable of a thermodynamical / statistical physical system [2]. In classical physics:

$$S_{tc} = k_B T \sum_{j=1}^{N} p_j \ln\left(\frac{1}{p_j}\right) \quad , \qquad (1)$$

where $p_j$ is the probability that the system is in the *j*-th microstate.

While thermal entropy measures allow an arbitrary additive constant in their value, for example depending on the resolution of counting microstates, the change of entropy during physical processes does not have any arbitrary component and it is objectively defined by measurable macroscopic physical quantities:

$$dS_t = \frac{dQ}{T} \quad , \qquad (2)$$

where *dS* and *dQ* are infinitesimally small changes of the entropy and heat dissipation at quasi-equilibrium conditions.

*2.2 Information (theoretic) entropy*

Information (theoretic) entropy [3] is a measure of the uncertainty of data in an information channel. Its value is determined by the message properties, the transmitter's parameters, the physics of the information channel and the properties of the receiver. When the information entropy is about a physical system, then the physical system forms the message and the transmitter and the measurement instrument forms the information channel and the receiver. In the case of binary data acquisition with *N* bit resolution, Shannon's information entropy is:

$$S_I = \sum_{j=1}^{N} \sum_{m=0}^{1} p_{j,m} \log_2\left(\frac{1}{p_{j,m}}\right) \qquad (3)$$

where $p_{j,m}$ stands for the probability of the *j*-th bit being in the bit value $m$ $(m \in \{0,1\})$. Equations 1 and 3 look very similar except the $k_B T$ multiplication factor in Equation 1 (providing thermal entropy unit) and the 2-base logarithm in Equation 3 (making a yes/no situation with probability 0.5 to have 1-bit of information). (Note, a similar "micro-state" approach, where the micro-state is the particular bit pattern in the *N*-long string would be more obviously identical to physics situation, and is extensively used for the private key in secure communications. However, for independent bits the two approaches produce identical results and our version is more suggestive to account for the energy dissipation, which results from single bit flips).

Similarly to thermal entropy, the resolution also represents an arbitrary additive constant in information entropy. However, there is a major difference between the two entropies. As we pointed out above, the changes of thermal entropy are objective. This statement is not necessarily true for the information entropy.

For example, imagine a message propagating in the channel. At a given bit resolution of the receiver, the information entropy is $S_I(1)$. If the bit resolution of the receiver is increased, the information entropy of the message will increase:

$$S_I(2) > S_I(1) . \tag{4}$$

(Note in practical situations with non-zero noise, when the increased bit resolution becomes fraction of the noise level then the impact of further increasing of the bit resolution on the information entropy saturates.)

When the message is received, it becomes a deterministically known record of data. For any deterministically known data set the information entropy is zero [4] because then the probabilities in Equation 3 are either zero or one. Thus the initial information entropy changes to zero

$$S_I(1) \to 0 \quad \text{and} \quad S_I(2) \to 0 \tag{5}$$

and during that process the entropy changes are different:

$$\Delta S_I(1) \neq \Delta S_I(2) . \tag{6}$$

In conclusion, even the change of the information entropy is subjective to the receiver, that is, to the measurement instrument, which is a striking difference compared to thermal entropy.

*2.3 Brillouin's and Landauer's principles*

Already the simple definitions above suggest that thermal entropy is an *objective* property of the physical system while the information entropy contains *subjective* elements that reflect on the characteristics of the measurement instrument. Yet, two famous claims, one by Brillouin and another one by Landauer-Bennett, seem to indicate a general interrelation between the two entropies.

Brillouin [5] introduced the *negentropy principle of information* based on thought experiments claiming that decreasing the information entropy by *N* bits by a measurement increases the thermal entropy in the system at least by $Nk_B \ln(2)$, where the $k_B$ "calibration coefficient" is the Boltzmann constant and the $\ln(2)$ factor stems from the different bases of the logarithm functions in Shannon's information entropy $(\log_2)$ and thermal entropy $(\ln)$. Brillouin's principle basically claims that the two entropies are additive, thus there is a "total-entropy" (information + thermal), which is exposed to the Second Law of Thermodynamics. In other words, it claims, that in a closed system, this "total entropy" cannot decrease, only increase [5] (see a similar picture in [6] in a different context):

$$\Delta S_t + k_B \Delta S_I \geq 0 , \tag{7}$$

where the equality $\Delta S_t + k_B \Delta S_I \simeq 0$ stands for the idealistic, "lossless" situation.

We will show in Section 4 that the real situation can be subtler than Brillouin's approach particularly when the PS and MI can be separated and/or when they have a different temperature. For example, the information entropy reduction about the PS may not cause entropy increase in the PS.

A decade later, first Landauer [7] and later Bennett (see, e.g. [8,9]) went beyond Brillouin and claimed *Landauer's principle* for erasure, that is, that *only the erasure of data* is exposed to Brillouin's principle. Furthermore they claim that reversible logic gates can do the rest of computation in a dissipation-free way. Since then, Landauer's principle has been refuted by several independent groups of authors in many

*This version is accepted for publication in the Journal of Computational Electronics, on July 17, 2017. (revision July 14, 2017). https://arxiv.org/abs/1706.01459*different ways [10-21] while in the majority of the literature its existence is taken granted. Our present paper also makes it clear that the Landauer principle is invalid.

In the rest of the paper, we illustrate the differences between the thermal and the information entropies by analyzing the situation of energy dissipation in simple measurement schemes designed to elucidate the differences between the two types of entropies.

## 3  Error versus energy dissipation at classical information processing

To prepare the evaluation of the energy dissipation in the various measurement schemes in Section 4, here we survey a few well-known-but-often-neglected facts about the minimum energy dissipation during changing the values of physical bits.

Measurement and data acquisition systems can be *analog* (continuum) or *digital* (binary, etc). Digital systems use thresholds to quantify the strength of the signal. Analog systems, where data with continuum values are processed without threshold elements, require higher energy dissipation than binary elements with equivalent performance due to the need to sustain a continuous power dissipation, see [20,21]. Thus, similarly to former efforts [5,22,23] we also focus on digital systems.

*3.1 Double potential well devices*

In most binary (two-state) gate operations, the fundamental lower limit of energy dissipation is given by the need to control a switching device that contains a potential barrier $E$ (as part of a single or double-well potential, see Figure 1), which is the essential component of switches; memories or their control; and it is also part of chemical reactions of data detection and storage such as in old-style photosensitive films. The essential role of the potential barrier is to operate as a threshold device for (binary) detection and/or the protection of the obtained data bit against corruption by the thermal excitations (noise). The energy dissipation stems from the fact that, when the state moves over the top of the energy barrier into the other well, the potential energy $E$ at the top must be dissipated to keep the state around the bottom of the well. This is because the switching of such a device is a kinetic transition and not a static transition. Any attempt to reuse the energy would require more energy dissipation than the gained-back energy because such a scheme would require having several new control steps; each of them would have the similar energy dissipation requirement as the original system.

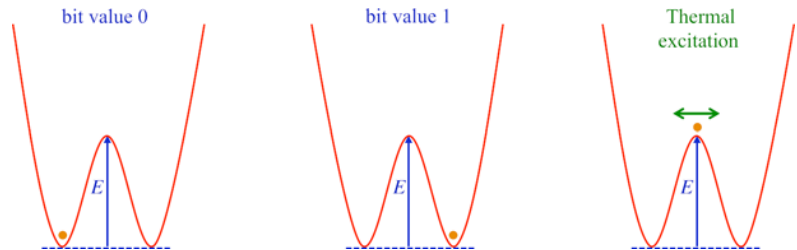

**Fig. 1.** Double potential well and bit errors via thermal excitation.

Brillouin [5], Alicki [22] and Kish [23], via different methods, arrived at the same result quantifying the minimum energy barrier height $E$ and the related energy dissipation $Q_{\min} = E$ in a symmetric double potential well to achieve the error probability $\varepsilon_0 < 0.5$ during flipping the bit value:

$$Q_{\min} = E \cong k_B T \ln\left(\frac{1}{\varepsilon_0}\right), \tag{8}$$

where $\varepsilon_0$ is the probability of finding the system in an erroneous state in the short observation time window limit, $t_w \leq \tau$, where $t_w$ is the observation time window and $\tau$ is the correlation time of thermal excitations.

For longer observation times $t_w$, each additional non-overlapping correlation time period $\tau$ represents a new independent "trial" for error generation resulting an increased error probability $\varepsilon_w = \varepsilon_0 \frac{t_w}{\tau}$ over $t_w$. Thus, in the small error limit, $\varepsilon_0 \ll 0.5\, \tau/t_w$, the generalized result for the minimum energy dissipation is [20,21]:

$$Q_{min} = E \cong k_B T \ln\left(\frac{1}{\varepsilon_w} \frac{t_w}{\tau}\right) \tag{9}$$

Equation 8 for $\varepsilon_0 \to 0.5$ yields the Szilard energy dissipation:

$$Q_{min} = k_B T \ln(2) \, , \tag{10}$$

which is also claimed by Brillouin's negentropy principle [5] as the lower limit of energy dissipation when measuring 1 bit of information and in Landauer's principle at erasing a bit of data.

Note:

(a) The same value is claimed by Landauer and Bennett [7-9] as the ultimate dissipation limit of erasing a bit however this claim is generally invalid [10-21].

(b) At the limit of $\varepsilon_0 \to 0.5$ the related memory content and operation of thermal demons are completely randomized which is rendering them useless. Thus only the relations for small error probabilities (Equations 8 and 9) have practical relevance.

## 4 Examples showing the differences between the two entropy measures

Note: The systems discussed below are supposed to be isolated from the impacts of environmental variations. Situations with cooled measurement instruments will also be discussed. These issues require making the following remarks:

∗ Cooling of the measurement instrument requires energy but that energy is not really related to the energy dissipation in the observed physical system. Of course, to keep the measurement instrument at the steady cold temperature, one must cool against the heat generated by the Brillouin principle within the MI. It is possible to calculate the minimum energy need for that (versus temperature: inverse Carnot cycle for cooling) but this issue is out of our main goal because this energy dissipation relates to the measurement instrument and it is irrelevant for the energy dissipation in the physical system (versus the info entropy change in the measurement instrument), which is the focus our current paper.

∗ Though the systems are supposed to be isolated from the environment that cannot really be true for the cooled measurement system. However, if an active temperature control is used then the environmental influence can be made negligibly small by sufficiently large loop-amplification of the control system and by using a stable reference voltage source.

*4.1 On classical physical measurements*

The experimental example in subsection 4.2 will feature two relevant characteristics of classical physical measurements, see Figure 3. Both factors can break Brillouin's negentropy principle:

(i) The temperature of the MI can be set to much lower than the temperature of the PS, that is, $T_m \ll T_s$. Thus, even though Equations 8-10 describe the bit-switching related heat dissipation remain relevant for the MI because of the need of at least a single switching operation per bit, this energy dissipation per bit within the MI can be much less than the value required by Brillouin's principle applied to the physical system, that is, $Q_{m,\min} \cong k_B T_m \ln\left(\frac{1}{\varepsilon_w}\frac{t_w}{\tau}\right) \ll k_B T_s \ln\left(\frac{1}{\varepsilon_w}\frac{t_w}{\tau}\right)$ .

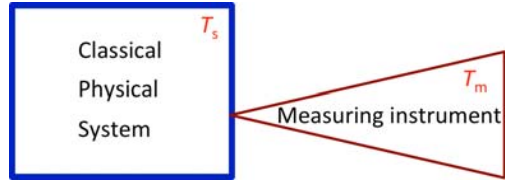

**Fig. 3.** Classical physical measurement. The measurement instrument can have so weak coupling to the physical system that the energy communication between them is approaching zero (see Figure 4 as example.)

(ii) In many classical physical systems, it is possible to design experiments that cause negligible disturbance to the physical system (PS) by the measuring instrument (MI). That means that, during measurement, the MI cause only an infinitesimal increase of thermal entropy (heat) *in the PS* while the reduction of the information entropy *in the MI* is significant. Therefore, the minimum energy dissipation $Q_{s,\min}$ caused by the perturbation by the MI in the PS can approach zero, $Q_{s,\min} \to 0$, and it can made negligible compared to the minimum energy dissipation in the measurement instrument $Q_{s,\min} \ll Q_{m,\min}$ even if Relations 8-10 hold (see the example in Section 4.2). In this case, the comparison of the changes of thermal entropies in the PS ($\Delta S_{st}$) and in the MS ($\Delta S_{mt}$) yields:

$$T_s \Delta S_{st} = Q_{s,\min} \ll Q_{m,\min} = T_m \Delta S_{mt} \qquad (11)$$

and

$$\Delta S_{st} = \frac{Q_{s,\min}}{T_s} \ll \frac{Q_{m,\min}}{T_s} = \frac{T_m \Delta S_{mt}}{T_s} \ll \Delta S_{mt} \qquad (12)$$

Assuming the idealistic, "lossless" situation of Brillouin's principle, $\Delta S_t + k_B \Delta S_I \simeq 0$ for the MI, we get:

$$\Delta S_{mt} \simeq -k_B \Delta S_I \qquad (13)$$

where $\Delta S_I$ is the change of information entropy in the MI about the state of the PS due to the measurement.

Thus, in accordance with Equations 12 and 13,

$$\Delta S_{st} \ll -k_B \Delta S_I \qquad (14)$$

which means that

$$\Delta S_{st} + k_B \Delta S_I < 0 \tag{15}$$

indicating the violation of Brillouin's generalized Second Law (Equation 7), the negentropy principle of information.

It is important to emphasize that both conditions, the $Q_{s,min} \ll Q_{m,min}$ and the $T_m \ll T_s$ situations, can independently break Brillouin's negentropy principle. Thus the break can happen even at identical temperatures or identical energy dissipations. Thus the Brillouin principle is generally invalid whenever the coupling between the PI and the MI is weak thus we can talk about two virtually separated systems.

On the other hand, by changing the relation between the temperatures and the generated heat by changing the physical parameters, it is possible to produce situations when Brillouin's inequality is satisfied but that does not change our assessment that no generally valid Second-Law-type relation exists involving the two types of entropies. In conclusion, Brillouin's negentropy principle is not a general law.

After these preparations, below we show some relevant experimental examples with practical parameters.

*4.2 A classical physical example: measuring 1 bit of information in thermal noise*

Suppose we have a parallel *RC* circuitry, which is the PS, and the task is to determine the sign of thermal noise voltage on the capacitor at time $t_1$, which is a 1 bit information provided the measurement is idealistically error free.

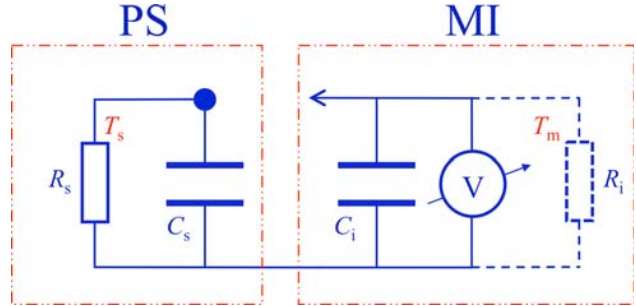

**Fig. 4.** Determination of the sign of the Johnson noise voltage (thermal noise) on a parallel RC circuit. The MI is briefly connected to the PS to take a sample of the thermal noise voltage on the capacitor $C_s$. The sample is held by the $C_i$ input capacitor. The voltmeter then measures and displays the result. Note, the measurement of the sampled voltage on $C_i$ can be done later thus the energy dissipation in the physical system and the information entropy change can be separated in time indicating their unconnected nature: the energy dissipation in the PS takes place even when the voltage measurement of the sample is skipped and the information entropy does not change, see Section 4.4.

The MI is a voltmeter and its input resistance $R_i$ can be passive or active (reactance) element. It satisfies $R \ll R_i$. The input capacitance of the voltmeter satisfies $C_i \ll C_s$. The temperature $T_s$ of the PS can be different from the effective noise temperature $T_m$ of the voltmeter.

First suppose that the $T_m \to 0$. When the voltmeter is connected, its input capacitance will extract

$$\Delta E = k_B T_s C_i / C_s \tag{16}$$

average energy from the system.

If $T_m = T_s$, then the extracted energy can be positive or negative and its expected value is zero but its rms value is similar:

$$\Delta E_{rms} = \left\langle \left(\Delta E\right)^2 \right\rangle^{0.5} \approx k_B T_s C_i / C_s . \tag{17}$$

According to standard knowledge about connecting charged capacitors, the same energy is dissipated in the wire upon connection. Thus, if we suppose that the dominant part of the connecting wire resistance is located in the PS, the energy dissipation $Q_s$ in the PS is:

$$\Delta S_{st} T_s = Q_s \approx k_B T_s C_i / C_s << k_B T_s , \tag{18}$$

that is,

$$\Delta S_{st} << k_B . \tag{19}$$

On the other hand, if the (internal) input noise of the voltmeter is sufficiently small, then the sign of the measured voltage can be determined by a small error probability $\varepsilon << 0.5$. For non-zero error probability, for a single bit of information, Equation 3 leads to Shannon's well-known reduced-information result [3]:

$$S_I = 1 + \varepsilon \log_2 \varepsilon + (1-\varepsilon) \log_2 (1-\varepsilon) < 1 \text{ bit} , \tag{20}$$

which converges to 1 bit for zero error probability:

$$\text{for } \varepsilon \to 0, \quad S_I \approx 1 \text{ bit} . \tag{21}$$

By the measurement of the sign of the voltage, this ~1 bit information entropy changes to zero thus the information entropy about the system changes by $\Delta S_m \approx -1$ bit .

Using Equation 19, we obtain:

$$\Delta S_{st} + k_B \Delta S_I \approx -k_B < 0 \tag{22}$$

which means the violation (Equation 15) is confirmed also in this particular measurement scheme indicating that Brillouin's negentropy principle on the change of information entropy *about* the PS and the related change of the thermal entropy *in* the PS is violated.

*4.3 A practical example with homogeneous temperature*

Here we show a practical example proving that the assumptions in Section 4.2 about the PS and MI are valid. Suppose that the physical system and the measurement instrument are at room temperature, that is, $T_s = T_m = 300 \text{K}$. The PS consists of standard commercially available components, $R_s = 4*10^5 \Omega$ and $C_s = 10^{-9} \text{F}$. The MS is also commercially available: the SR 560 preamplifier and an arbitrary AC digital voltmeter connected to its output. The SR 560 has $C_i = 10^{-11} \text{F}$ and $R_i = 10^8 \Omega$ with equivalent input noise data given below.

The effective thermal noise $U_c$ to measure on the $C_s$ is [24]:

$$U_c = \sqrt{\frac{k_B T_s}{C_s}} \approx 2*10^{-6} \text{ V} \tag{23}$$

with Lorentzian power density spectrum and -3 dB cut-off frequency:

$$f_c = (2\pi R_s C_s)^{-1} \approx 400 \text{ Hz} \quad . \tag{24}$$

To follow the variations of the measured voltage with a high-fidelity, let us set the cut-off frequency $f_m$ of the preamplifier-voltmeter system to a ten-fold higher frequency (which is 5 times higher than the Nyquist sampling frequency for this bandwidth):

$$f_m = 4000 \text{ Hz} \quad . \tag{25}$$

The equivalent input voltage noise spectrum of the SR 560 preamp is $S_u \approx 1.6*10^{-17}$ V$^2$/Hz and its equivalent current noise spectrum is $S_i \approx 10^{-28}$ A$^2$/Hz thus the total equivalent input noise spectrum interfering with the measured voltage is:

$$S_{tot} = S_u + R_s^2 S_i = 3.2*10^{-17} \text{ V}^2/\text{Hz} \quad , \tag{26}$$

which, in the given bandwidth, results in the following total effective noise interference:

$$U_{mn} = \sqrt{f_m S_{tot}} \approx 3.6*10^7 \text{ V} \tag{27}$$

The bit error probability of the sign determination will be:

$$\varepsilon \approx \frac{U_{mn}}{U_c} \approx 0.18 \tag{28}$$

According to Equation 18, this error probability provides:

$$S_I \approx 0.32 \text{ bit} \tag{29}$$

information for before measurement. Therefore, by the measurement, the change of information entropy is:

$$\Delta S_I \approx -0.32 \text{ bit} \quad . \tag{30}$$

At the same time, the energy dissipation in the system is (see Equation 18):

$$\Delta S_{st} T_s = Q_s \leq k_B T_s C_i / C_s = 0.01 k_B T_s \quad , \tag{31}$$

corresponding to

$$\Delta S_{st} \leq 0.01 k_B \quad , \tag{32}$$

where the equality is for the most pessimistic case when all the dissipating wire located in the PS. Thus in the highest energy dissipation case, Equations 22 and 32 yield:

$$\Delta S_{st} + k_B \Delta S_I = 0.01 k_B - 0.32 k_B = -0.31 k_B \quad . \tag{33}$$

In conclusion, Brillouin's negentropy principle is violated for the measured physical system when we can separate the measurement instrument, such as in this case.

*4.4 Separation of the thermal and information entropy changes in space and time*

It is important to recognize that that reason why the Brillouin's (negentropy) principle is broken in the above examples is that we can separate the thermal entropy changes in the PS and in the MI. For the thermal entropy in the MI, Brillouin's principle is valid because at such a measurement at least a single switch must alternate and Equations 8-10 guarantee the necessary energy dissipation to satisfy the negentropy principle (Equation 7).

It is even more interesting that the thermal entropy change in the PS and the information entropy change in the MI can be separated also in time, see Figure 4. In *delayed-evaluation* scheme, the measurement of the sampled voltage on $C_i$ can be done later thus the energy dissipation in the physical system and the information entropy change can be separated in time indicating their unconnected nature. The only physical condition for that is the proper selection of the time constant $R_i C_i$ :

$$t_d \ll R_i C_i \quad . \tag{34}$$

(i) It is important to recognize that, in this case, the energy dissipation in the measured physical system happens earlier than the change of the information entropy about the system.

(ii) The energy dissipation in the PS takes place even when the voltage measurement of the sample is skipped and the information entropy does not change.

In conclusion, there is no causal relation between the change of information entropy about the system and the energy dissipation in the system related with that measurement.

**5 Data erasure in memories**

In a memory, the data are deterministic and generally known. If they are not known, they can be measured before and independently of erasure. Therefore during *erasure-by-resetting*, the information entropy in the memory is zero before and after the erasure. Thus the information entropy does not change and there is no way to interrelate the changes between the information entropy and the thermal entropy.

The situation of *erasure-by-randomization* is somewhat different. Here we show our earlier argument [18], which is the simplest evidence why Landauer's principle of erasure dissipation is invalid. The natural process that leads to the dissipation-free erasure, thus proving Landauer's theorem invalid, is thermalization in a symmetric double-well potential system, such as in a magnetic memory; see Figure 1. When such a system is kept untouched for time $t_e$ that is much longer than the thermal relaxation time

$$t_e \gg \tau \exp\left(\frac{E}{k_B T}\right) , \tag{35}$$

the memory cells get thermalized and the probability of finding them in the 1 and 0 states exponentially converges to 0.5. This erasure-by-randomization process occurs without energy dissipation because equilibrium thermal fluctuations are utilized for erasure. Of course, such a process may take thousands of years, but there is no time restriction in the Landauer's principle. The existence of this phenomenon proves that no energy dissipation is required for information erasure thus Landauer's way to interrelate information and thermal entropies is also invalid.

**6 Information entropy and the Third Law of Thermodynamics**

The Third Law of Thermodynamics states [2] that the thermal entropy converges to zero when the zero absolute temperature limit is approached (at least in crystalline structures). This property of thermal entropy sharply differs from the behavior of information entropy because there are situations where the information entropy does not follow the Third Law. The simplest example is the thermal noise voltage of resistors when the temperature of the resistor approaches zero. In this case, the experimentally observable noise voltage and the related information entropy depends on the type of the measurement system:

x) If the measurement is done by a linear, amplitude/phase sensitive amplifier then the observed thermal noise is *not zero and it is divergent for high frequencies* [25].

y) If the measurement is done by an antenna that transforms the thermal noise voltage and current of the resistor to radiation, and this radiation is measured by a photocell, then the evaluated thermal noise of the resistor is *zero*, in accordance with Planck's black body radiation formula [26].

z) If the measurement is done by measuring the force between the plates of the $C_s$ shunt capacitor in the cicuitry in Figure 4, then the observed thermal noise is again *zero*, otherwise the Second Law is violated [27].

In conclusion, these considerations also verify that the information entropy is not an objective quantity and its value depends even on the type of the measurement device. Furthermore, at situation (x), the information entropy violates the Third Law because the information entropy is not zero due to the measured non-zero noise. On the other hand at situations (y) and (z) the Third Law is satisfied because the zero thermal noise voltage indicates a deterministic measurement result.

## 7 Conclusions

In this paper, we have shown that the two major efforts to find general physical principles that interrelate changes in the information entropy and the thermal entropy in physical systems are invalid. Particularly:

i) In general, the information entropy and its changes contain a component that is subjective to the measurement instrument, while the changes of thermal entropy can be stated objectively.

ii) Brillouin's negentropy principle of information (the expanded formulation of the Second Law by Equation 7), is *invalid as a general rule* because violations can also occur in a physical system provided the temperature of the measurement system is less than that of the measured physical system.

iii) In the case of homogeneous temperatures, it can be *seemingly valid* if the measurement system is integrated with the measured physical system, or if not, it can be valid *within the measurement system alone*. However, in classical physical situations, the measurement system and the measured physical system, as well as the change of information entropy and the related change of thermal entropy can be separated in space and time leading to the break of Brillouin's negentropy principle.

iv) The information entropy can increase without triggering any change of the thermal entropy indicating that information erasure does not necessarily require energy dissipation.

v) *There is no case where Landauer's principle of erasure dissipation is even seemingly valid* because erasing of the memory does not yield change in the information entropy.

vi) The information entropy can violate the Third Law of Thermodynamics.